%% file: main.tex
\documentclass[aps,prab,twocolumn,10pt,groupedaddress]{revtex4-2}
\usepackage{tikz}
\usepackage{graphicx}
\usepackage{url}
\usepackage{fancyvrb}
\usepackage{color}
\usepackage{tikz}
\usepackage{hyperref}
\usepackage{float}

\usetikzlibrary{shapes,arrows,positioning,shapes.geometric}
\usepackage{charter}
\usepackage{siunitx}
\usepackage{amssymb, amsmath}
\usepackage{graphicx}
\usepackage{color}
\usepackage{afterpage}

\usepackage{struktex}

\usepackage{xspace}

\newcommand{\hp}{\texorpdfstring{\ensuremath{\mathrm{H}^+}\xspace}{H+}}
\newcommand{\htp}{\texorpdfstring{\ensuremath{\mathrm{H}_2^+}\xspace}{H2+}}
\newcommand{\hthp}{\texorpdfstring{\ensuremath{\mathrm{H}_3^+}\xspace}{H3+}}
\newcommand{\hyd}{\texorpdfstring{\ensuremath{\mathrm{H}_2}\xspace}{H2}}

\newcommand{\BE}[0]{\begin{equation}}
\newcommand{\EE}[0]{\end{equation}}
\newcommand{\BEA}[0]{\begin{eqnarray}}
\newcommand{\EEA}[0]{\end{eqnarray}}

\mathchardef\mhyphen="2D

\newcommand{\figref}[1]{FIG.~\ref{#1}}
\newcommand{\tabref}[1]{TAB.~\ref{#1}}

\begin{document}


\title{High current \htp beams from a filament-driven multicusp ion source}


\author{Daniel Winklehner}
\email[]{winklehn@mit.edu}
\author{Janet M. Conrad}
\author{Joseph Smolsky}
\author{Loyd Waites}
\affiliation{Massachusetts Institute of Technology, 77 Massachusetts Ave, Cambridge, MA, USA}


\date{\today}

\begin{abstract}
Recently, the use of \htp ions instead of protons to overcome space charge challenges in compact cyclotrons has received much attention. This technique has the potential to increase the available beam current from compact cyclotrons by an order of magnitude, paving the way for applications in energy research, medical isotope production, and particle physics, e.g. a decisive search for sterile neutrinos through the IsoDAR experiment. For IsoDAR we go a step beyond just using \htp and add pre-bunching through a Radio-Frequency Quadrupole (RFQ) embedded in the cyclotron yoke. This puts beam purity and beam quality constraints on the ion source that no published ion source has simultaneously demonstrated so far. Here, we report results from a new multicusp ion source (MIST-1) that produces the world’s highest steady-state current of \htp from this type of ion source (1~mA), with exceptionally low emittance (0.05~$\pi$-mm-mrad, RMS, normalized) and high purity (80\% \htp). This result shows the feasibility of using a multicusp ion source for IsoDAR and the RFQ direct injection prototype, and paves the way to record breaking continuous wave (cw) beam currents of 5~mA \htp (equivalent to 10~mA protons) from compact cyclotrons, ideal for underground installation. This represents a significant advance, with impact on neutrino physics specifically and high power cyclotron design in general.
\end{abstract}


\maketitle


\input{Sec1_Introduction}

\input{Sec2_Setup}

\input{Sec3_Simulations}

\input{Sec4_Measurement}

\input{Sec5_Conclusion}

\begin{acknowledgments}
This work was supported by NSF grants PHY-1505858 and 
PHY-1626069, as well as funding from the Bose Foundation. 
The authors are very thankful for the support of the MIT Central Machine shop, the 
MIT Plasma Science and Fusion Center (PSFC), and the University of Huddersfield 
for support with machining, lab space and utilities, and equipment, respectively.
Furthermore, the authors would like to thank Jose Alonso and William Barletta for
fruitful discussions.
\end{acknowledgments}

\bibliographystyle{apsrev4-2} 
\bibliography{MIST1_Bibliography}

\end{document}

%% file: Sec1_Introduction.tex
\section{Introduction}


Compact cyclotrons that can produce 10~mA proton beams with energy $>50$ MeV, 
and that can be mass-produced, would provide a powerful tool for basic 
\cite{bungau:isodar1, abs:isodar_cdr1, winklehner:nima} and 
applied science \cite{alonso:isotope, schmor:isotopes, ishi:adsr}.  
This is an order of magnitude higher current than is available from on-market cyclotrons.   
It has recently been shown that the solution to many of the challenges of achieving this goal 
lies in accelerating 5 mA of \htp ions \cite{winklehner:nima}. 
In this paper we report on an important first step in the development of such an 
accelerator: The design of the MIST-1 ion source and its commissioning at reduced 
power, which nonetheless already yielded record \htp currents. 

\begin{figure}[t!]
    \includegraphics[width=0.9\columnwidth]{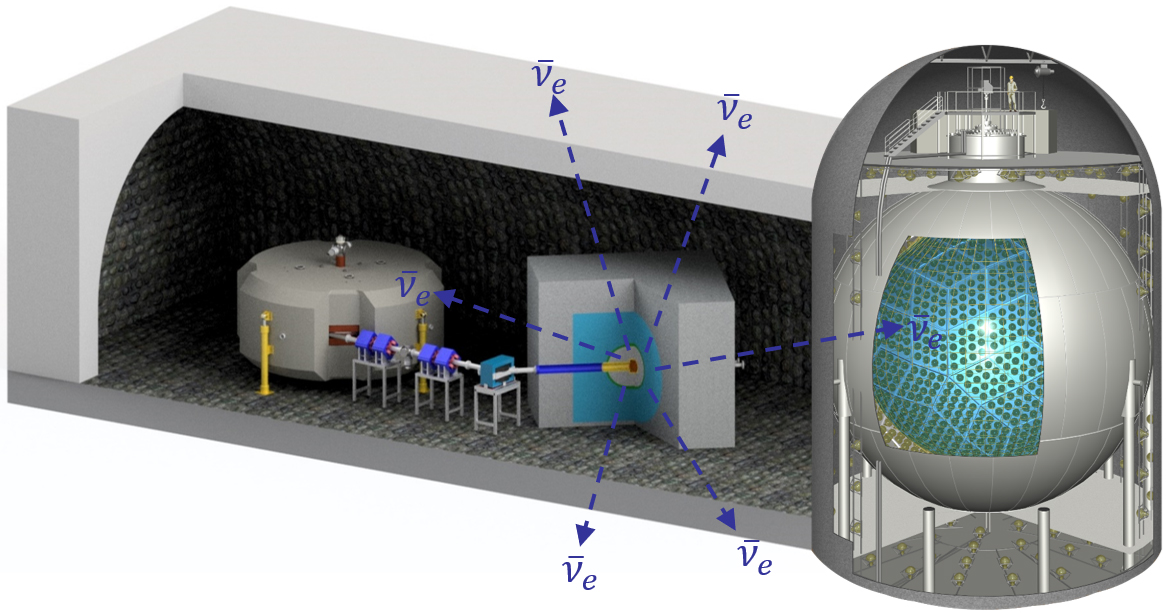}
    \caption{Artists rendition of the IsoDAR experiment paired with the KamLAND 
             detector at Kamioka. From left to right: The cyclotron
             (ion source on top),
             generating a 60 MeV/amu \htp beam, the medium energy beam transport line,
             the neutrino production target \cite{bungau:target1, bungau:target2}, 
             and the KamLAND detector \cite{kunio:kamland}. 
             \label{fig:isodar_cave}}
\end{figure}

MIST-1 is an integral part of the IsoDAR project, and the requirements of this 
experiment have set the ion source specifications.
IsoDAR is planned as a definitive search for anti-electron-flavor 
disappearance due to eV scale sterile neutrinos, a beyond Standard Model particle, 
as described in Refs. \cite{bungau:isodar1, abs:isodar_cdr1, winklehner:nima}. 
The novelty of IsoDAR is in constructing an intense low energy anti-neutrino 
source near a kiloton scale neutrino detector, KamLAND \cite{kunio:kamland}, 
that is located underground, to reduce backgrounds (see Fig, \figref{fig:isodar_cave}).
The anti-neutrinos are produced by a 10~mA, 60~MeV continuous wave (cw) proton beam
impinging on a $^8$Be target surrounded by $^7$Li.
To save on space and costs, IsoDAR will use a compact cyclotron as a driver 
instead of a linear accelerator.
Commercially available cyclotrons in this energy range typically have maximum 
proton beam currents around 1~mA, with space charge the primary limiting factor. 
In order to reduce space charge effects, IsoDAR will accelerate \htp instead of 
protons or H$^-$ (note that \hthp was also investigated, but the higher rigidity
would make the cyclotron larger and more expensive).
With this, and additional innovations described in Ref. \cite{winklehner:nima}
(axial injection through an RFQ, high accelerating gradient, vortex-motion),
the IsoDAR cyclotron will accelerate 5~mA of \htp to 60~MeV/amu, which, after 
charge-stripping, yields 10~mA of protons at 60~MeV on target.

The ion source requirements for RFQ direct axial injection into the 
compact cyclotron are:
\begin{enumerate}
\setlength\itemsep{0.1em}
    \item Low emittance ($<0.1$ $\pi$-mm-mrad, rms, norm.)
    \item Low contamination ($>80$\,\% \htp fraction)
    \item High current (10~mA of \htp) 
\end{enumerate}
These stem from the desire to keep the system compact, and utilize the RFQ's ability
to separate by mass in addition to
its highly efficient bunching and pre-acceleration.
In \figref{fig:isodar_cave}, the entire low energy beam transport is contained 
within the 6-way cross on top of the cyclotron.

The status of high current \htp and \hthp sources was recently reviewed 
in \cite{wu:hydrogen}, 
which lists 2.45~GHz Electron Cyclotron Resonance 
(ECR) ion sources \cite{miracoli:vis1, xu:ecr1} and filament-driven discharge 
sources, with the plasma either confined by a solenoid \cite{schweizer:franz1} or by a 
multicusp field \cite{ehlers:multicusp1}.
For IsoDAR, the 2.45\,GHz ECR versatile ion source (VIS) 
\cite{castro:vis1, castro:vis2} was first tested by the
MIT group in collaboration with INFN-LNS and Best Cyclotron Systems, Inc. 
While the performance for protons was excellent, the finding was that the 
maximum \htp fraction was limited to 50\,\% at the required total
beam currents \cite{alonso:vis1, winklehner:bcs_tests}. 
In \cite{wu:hydrogen} similar fractions are reported.
In Ref. \cite{joshi:franz1}, a filament discharge source with solenoid confinement 
was presented that showed $>91$\,\% \htp fraction and 2.84~mA of \htp, however,
the energy spread was on the order of 120~eV (reported for a He$^+$ beam).
In Ref. \cite{ehlers:multicusp1}, a multicusp ion source, developed at LBNL,
capable of producing extractable total current densities of 
50\,mA/cm$^2$, with up to 80\,\% of the beam being H$_2^+$ ions was presented.
Combined with typically low emittances (e.g. $<0.04$~$\pi$-mm-mrad 
for Ar+ \cite{wutte:multicusp}), 
low energy spread ($<2$~eV \cite{lee_multicusp}), and ease of operation,
this type was deemed the best choice for a new source.
It is important to note that the results of the LBNL source were for sub-mm diameter 
extraction apertures and pulsed beams.
To fulfil the IsoDAR requirements, MIST-1 was designed to deliver 10~mA of \htp 
beam current in DC mode, requiring significant design changes and cooling upgrades.

\begin{figure}[t!]
	 \center{\includegraphics[width=0.90\columnwidth]
        {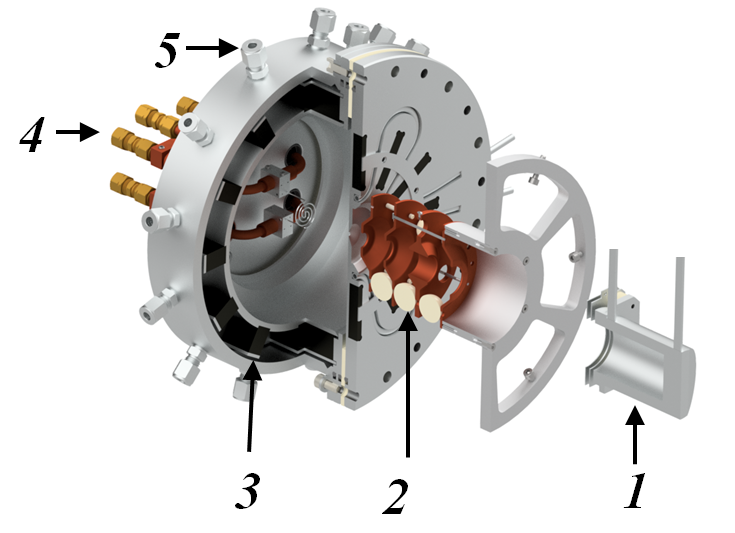}}
	    \caption{\label{fig:source}3D CAD rendering of MIST-1 ion source with extraction system and Faraday cup 1 (FC1). 
	    Inner parts are labeled: 1. Faraday cup, 2. Extraction System,
	    3. Permanent magnets (Sm$_2$Co$_{17}$),  
	    4. Water cooled Filament feedthroughs, 5. Water cooling fittings. The backplate also has a gas inlet, through which the hydrogen gas may enter into the source.}
\end{figure}

In this early commissioning phase, MIST-1 has already set a new record
for this type of ion source: a 1.08 mA DC \htp beam. 
At full power, MIST-1 must deliver a 10~mA DC \htp beam to the RFQ and cyclotron.
With a demonstrated plasma density of 11~mA/cm$^2$ and doubling the aperture
size from 4~mm to 8~mm diameter, the scaled \htp current of 4~mA \htp, when run through
our detailed cyclotron simulations, would already allow building a compact cyclotron
that can deliver 4 times the current of a commercial system. Additional 
development is ongoing that will bring us to the nominal current of 10~mA \htp.
In addition to the beam currents, 
we show very low emittances of 0.05 $\pi$-mm-mrad (1-rms, normalized), which
are in good agreement with simulations.
Section \ref{sec:setup} contains a detailed description of the source and diagnostic setup. 
Section \ref{sec:sims} covers simulations of extraction system and test beam line. 
Measurements are in Section \ref{sec:measurements}.

%% file: Sec2_Setup.tex
\section{Experimental Setup \label{sec:setup}}
\subsection{Ion Source}
\begin{figure}[t!]
	 \center{ \includegraphics[width=1.0\columnwidth]
        {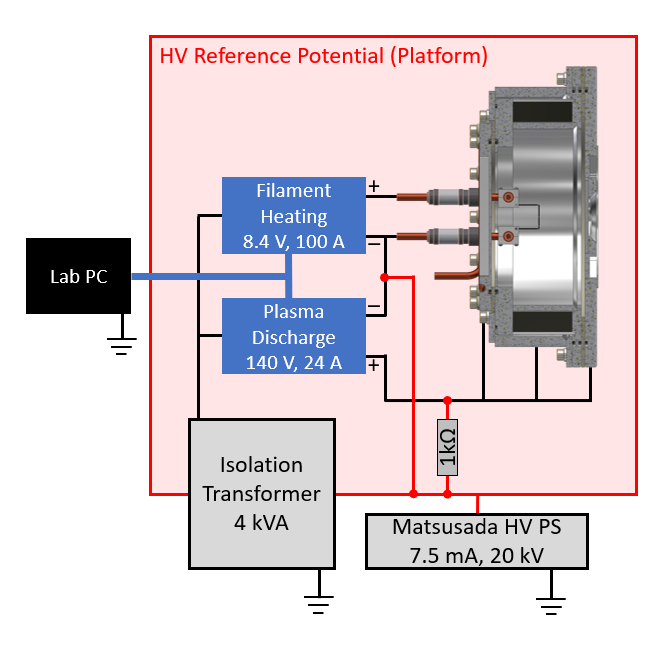}}
	    \caption{\label{fig:wiring_diagram} The wiring schematic of the ion source. Red is the high voltage reference potential, blue are the data cables, and black are power cables. The power supplies are computer-controlled via an optical USB extender cable. The source back plate, body, and front plate can all be held at different potentials (they are separated by insulator rings). During
	    the measurements presented here, they were held at the same potential.}
\end{figure}
The MIST-1 ion source is a filament-driven multicusp ion source and is described in detail in Refs. \cite{abs:isodar_cdr1, axani:mist1, winklehner:nima}. A CAD rendering is shown in
\figref{fig:source}, and a cross section view in \figref{fig:wiring_diagram} (right).
A stainless-steel chamber with samarium-cobalt permanent magnets creates a multicusp field that confines the plasma around the extraction aperture. Hydrogen enters the source through the backplate. The hydrogen undergoes electron-impact-ionization by electrons from the tungsten filament, which are being accelerated towards the anode. The filament is a tungsten alloy, mixed with small amounts of copper and nickel for corrosion resistance. The ions are then extracted through a hole in the center of the front-plate. The source is on a high voltage platform that can be lifted to a maximum potential of 20~kV. Ions leaving the source are focused and accelerated in the extraction system. The ion source parameters described here are summarized \tabref{tab:source}

The amount of hydrogen entering the source is controlled by a mass flow controller (MFC), (MKS Instruments, Model: GV50A, with 5~sccm full range).
The MFC is controlled via a RS485/USB with optical interface connected to 
our control system PC.

The filament is connected to two power supplies (see \figref{fig:wiring_diagram}): 
The filament heating power supply (8.4~V, 300~A max), to raise the temperature of the 
filament for thermionic electron emission; and the plasma discharge power supply 
(140~V, 24~A max), to maintain a potential difference between the filament (cathode) 
and the source body (anode), facilitating the discharge.
The source body, back plate, and front plate are all electrically insulated. 
This will allow each component to be maintained at a different potential in 
order to test the effects of varying the anode arrangements in future studies.
For this study, all components are maintained at the same potential.
The filament emits electrons that ionize the gas to form a plasma. 
\htp is a fragile ion that easily recombines to \hthp or dissociates
into protons through further collisions in the plasma. Its mean free path
is short at 1-2 cm \cite{axani:mist1, ehlers:multicusp1}. 
MIST-1 accounts for this with a short source body (low aspect ratio), producing 
the bulk of \htp within a few centimeters of the extraction aperture, not giving 
the ions enough time to undergo a significant number of collisions before 
drifting out of the source.

\begin{figure*}[t!]
        \includegraphics[width=0.8\textwidth]{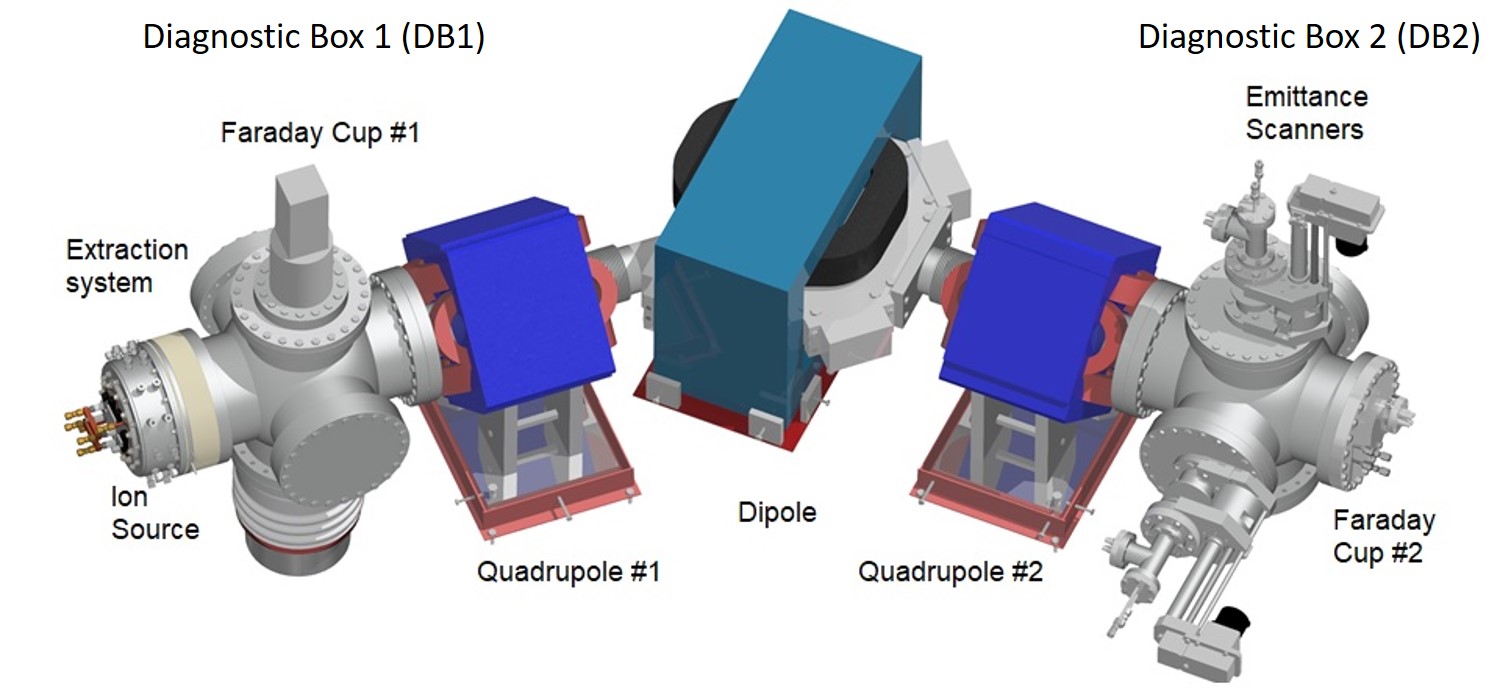}
        \caption{\label{fig:lebt_layout} 
        CAD model of the LEBT and diagnostic system. Starting with multiple species coming from the ion source, the species are then focused in the beamline by quadrupole magnets, and separated by the dipole magnet. Their currents and emittances can then be measured by the Faraday cup and emittance scanners in DB2.}
 \end{figure*}
The size, shape and position of the filament have an effect on the properties of the plasma. We have found that using a filament that is less than 0.8 mm thick does not provide sufficient electrons to generate a dense plasma. A systematic study for optimal filament shape, thickness, and position is forthcoming. The specific parameters of each filament used in the studies presented here will be described in the measurement section.

 \begin{table}[t!]
    \centering
    \caption{MIST-1 ion source parameters. \label{tab:source}}
    \begin{tabular}{ll}
        \hline
        \textbf{Parameter} & \textbf{Value (nominal)}\\
        \hline\hline
        Plasma chamber length & 6.5 cm\\
        Plasma chamber diameter & 15 cm \\
        Permanent magnet material & Sm$_2$Co$_{17}$\\
        Permanent magnet strength & 1.05 T on surface\\
        Front plate magnets & 12 bars (star shape)\\
        Radial magnets & 12 bars \\
        Back plate magnets & 4 rows of magnets, 6 bars tot.\\
        Front plate cooling & embedded steel tube  \\ 
        Back plate cooling & embedded copper pipe \\
        Chamber cooling & water jacket\\
        Water flow (both) & (1.5 l/min)\\
        Filament feedthrough cooling & water cooled \\
        Filament material & W mixed with Cu and Ni\\
        Filament diameter & $\approx 0.8 $ mm  \\
        Discharge voltage & max. 180 V\\
        Discharge current & max. 24 A \\
        Filament heating voltage & max. 8 V\\
        Filament heating current & max. 100 A\\
        \hline
    \end{tabular}
\end{table}

The ions drift out of the source through the extraction aperture where they are shaped and accelerated by a tetrode extraction system. The extraction system is a series of four copper electrodes that shape the beam upon leaving the ion source (see \figref{fig:ibsimu}) through static voltages applied by Matsusada power supplies AU-20P7 and AU-20N7, for positive and negative voltages, respectively. The electrodes following the source plate are the screening electrode, or ``puller'' (typically kept at a low, negative voltage of -2~kV) and the einzel lens. The puller prevents electrons from the beam line from streaming back into the ion source, and increases the electric field that shapes the meniscus. The einzel lens is made up of a total of three electrodes. The outer ones are grounded and the larger central piece is adjustable up to 20~kV. This feature is mostly unused in the presented measurements.
The electrodes are aligned via the compression of several ceramic balls. The extraction system is modeled using the IBSimu code \cite{kalvas:ibsimu}, discussed in Section \ref{sec:sims}.

\subsection{Low Energy Beam Transport and Diagnostics}
\begin{table}[!b]
    \centering
    \caption{\label{tab:dipole_parameters} 
             Parameters for the beam line magnets.}
    \begin{tabular}{llll}
        \hline
        \textbf{Dipole Parameter} & \textbf{Value} & \textbf{Quad Parameter} & \textbf{Value}\\
        \hline\hline
            Manufacturer & Bruker & Manufacturer & BNIP \\
            $\mathrm{B}_\mathrm{max}$ (center) & 0.7~T & 
            $\mathrm{B}_\mathrm{max}$ (pole tip) & 0.085~T\\
            $\mathrm{I}_\mathrm{max}$ & 125~A & $\mathrm{I}_\mathrm{max}$ & 200~A\\
            $\mathrm{U}_\mathrm{max}$ & 47~V & $\mathrm{U}_\mathrm{max}$ & 2~V\\
            Bending radius & 300~mm \hspace{5pt} & Pole tip radius & 37.5~mm\\
            Pole gap & 75~mm & Aperture diameter & 90~mm\\
        \hline
    \end{tabular}
\end{table}
The ion source extraction system is followed by a low energy beam transport line (LEBT) used for beam diagnostics (see \figref{fig:lebt_layout}). The LEBT consists of three electromagnets: 2 quadrupoles and 1 dipole.  The arrangement of the magnets, beamline, and diagnostics 6-way-crosses (designated as DB1 and DB2 in the text below) are shown in \figref{fig:lebt_layout}. Parameters for the magnets are listed in \tabref{tab:dipole_parameters}.
Both quadrupoles can be used for vertical focusing of the beam, however only the first quadrupole is used for the studies presented here. When a mass spectrum is being taken, the quadrupole current is increased in a constant ratio with the dipole current. The dipole is used for horizontal focusing and ion species separation.  Following the second quadrupole magnet is the analysis 6-way-cross (DB2), which contains a second Faraday cup and two Allison scanners with perpendicular axes.

\begin{figure}[b!]
        \includegraphics[width=.4\textwidth]{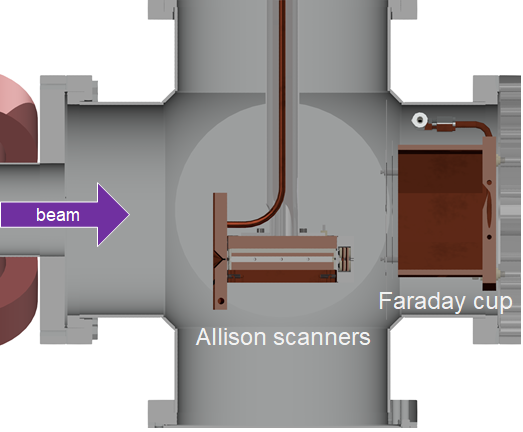}
        \caption{\label{fig:diagnostic_box} 
        CAD rendering of DB2, housing the Allison scanners (vertical scanner shown) and FC2. The Allison scanners can be retracted so that beam current can be measured in FC2
        to obtain mass spectra.}
 \end{figure}

 The Faraday cup (FC1) in DB1, which follows the extraction system, is used to measure the total current coming from the source without separation of species. A second Faraday cup (FC2) is located at the end of the beamline to measure the relative species fractions after separation by the dipole (see \figref{fig:diagnostic_box}).
 Each Faraday cup is equipped with a negative suppression electrode to prevent secondary electrons from escaping, which would artificially increase the measured positive ion current. The necessary suppression voltage for our Faraday cups has been experimentally determined to be -350 V.
 
To measure horizontal and vertical emittance (separately), 
DB2 houses two Allison-type electrostatic emittance scanners \cite{allison:emittance}.
A model for the identical Allison scanners, designed and constructed at MIT, is shown in Figure \ref{fig:diagnostic_box}.  They are equipped with a water-cooled copper front plate and are able to handle proton beam currents of up to 50~mA and 80~keV.  Further details are provided in Ref. \cite{corona2018emittance}.

\subsection{Measurement Parameters}
Systematic studies during commissioning of this source are done
varying the following parameters:
\begin{itemize}
\setlength\itemsep{0.1em}
\item Discharge voltage,
\item Filament heating current,
\item \hyd gas flow into source,
\item Filament shape and position.
\end{itemize}
Beam line parameters like source high voltage potential, einzel lens setting and
quadrupole settings are then adjusted for each data point to give the best 
transmission to FC2.

The filament discharge current strongly affects the plasma density. 
In order for the discharge current to be kept constant and stable, 
the heating current can be used to compensate for the changes in the plasma
(e.g. temperature of the filament, pressure fluctuations, etc.). 
This is controlled by a  proportional–integral–derivative control-loop, or ``PID loop'' \cite{PIDloop}.
As the filament heating and discharge voltage are varied, the total current is measured in FC1.  The platform and puller voltages are adjusted to keep the beam focused into FC1, which is monitored by measuring the current on its front plate.  When taking an 
ion mass spectrum, the dipole current is varied while current measurements are taken in FC2.  The first quadrupole's current is set proportional to the dipole current, for vertical focusing.
FC2 currents are then plotted as a function of dipole field (see Sec. IV). 
As ions with different mass-to-charge ratio have different magnetic rigidity, a
different dipole field strength is necessary to transport them to FC2, thus
leading to individual peaks in the spectrum. The magnets 
are controlled, and mass spectra are recorded, via an automated LabVIEW program 
run on the PC.

%% file: Sec3_Simulations.tex
\section{Simulations \label{sec:sims}}
\begin{figure}[b!]
    \includegraphics[width=1.0\columnwidth]{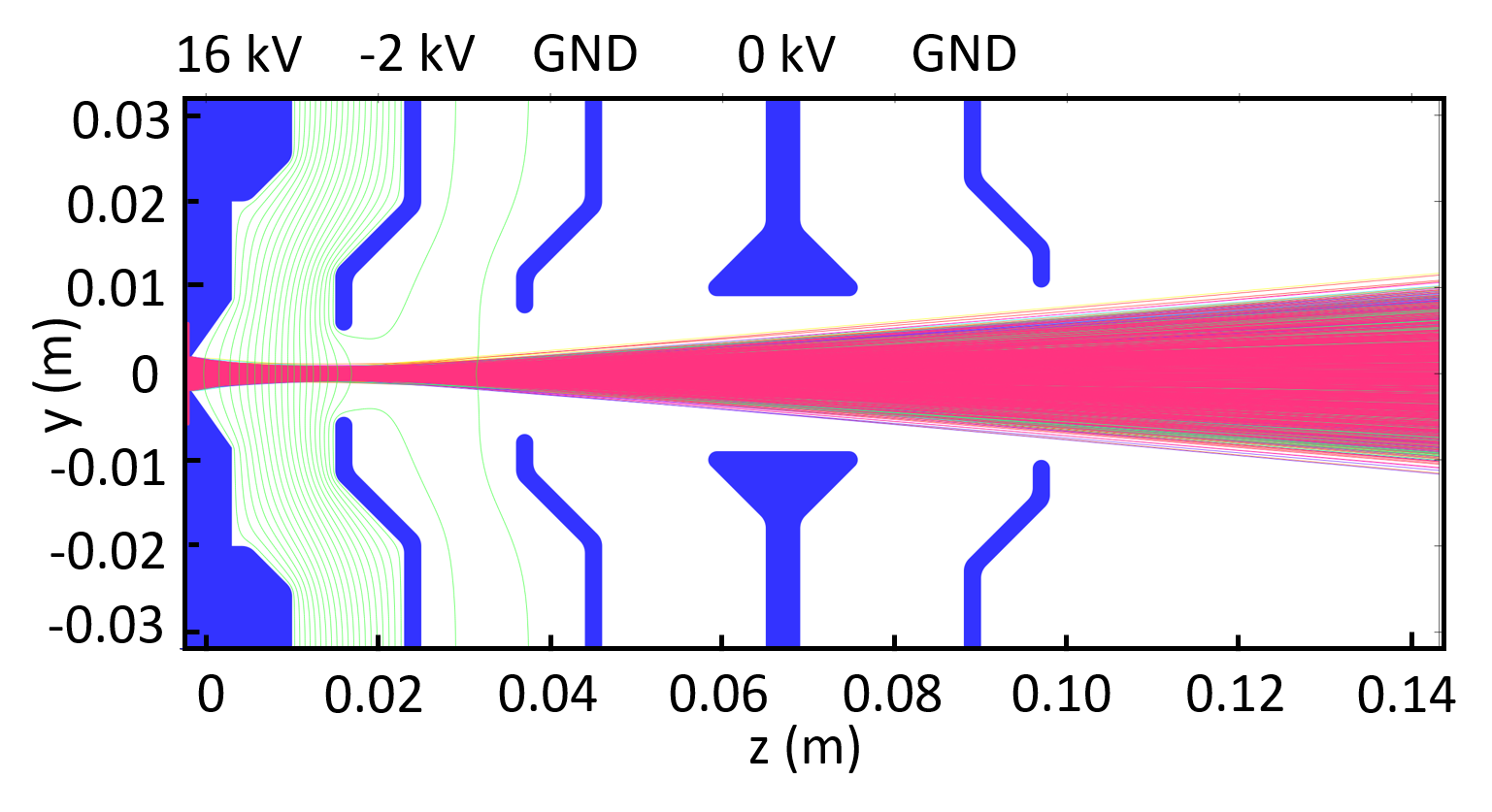}
    \caption{IBSimu simulation of the low energy extraction system. The electrodes are in blue, and the equipotential lines are in green. The electrode voltages are shown at the top.  This simulation includes all species listed in the text.\label{fig:ibsimu}}
\end{figure}
 \begin{figure*}[t!]
    \includegraphics[width=\textwidth]{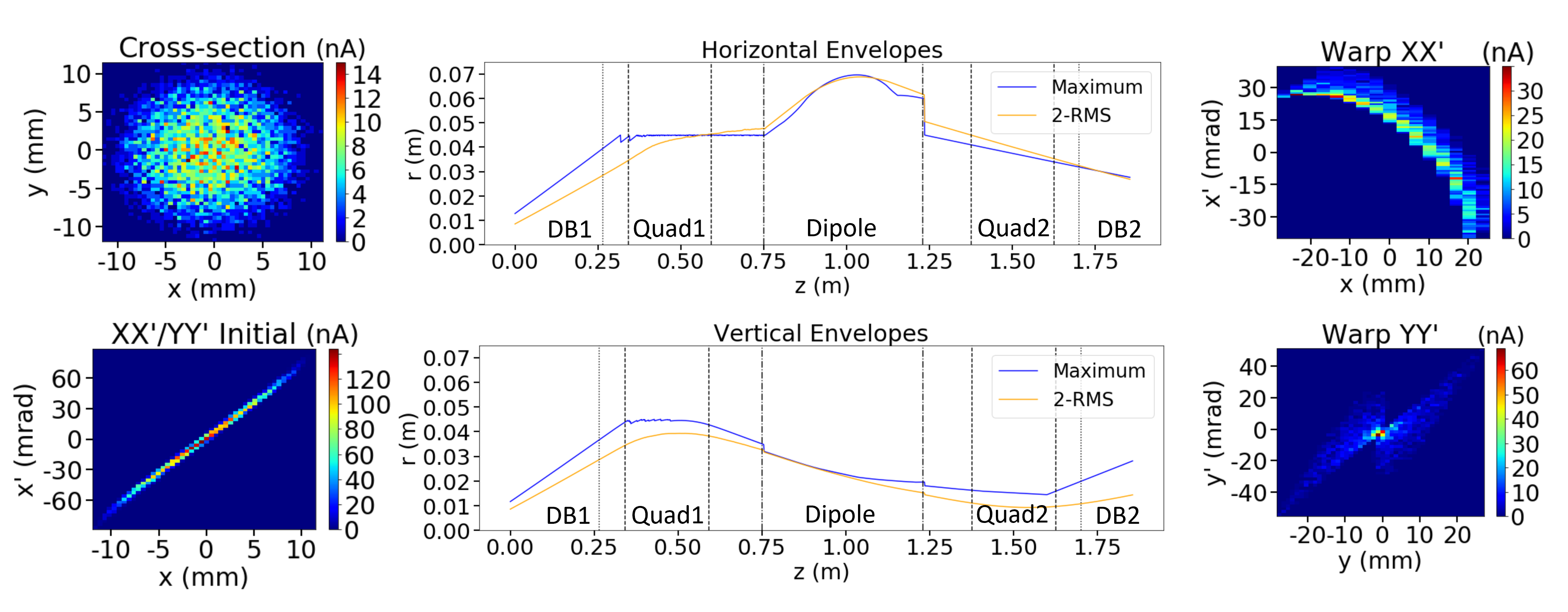}
    \caption{Example simulation of test beam line using Warp.  The simulation starts 14 cm from the extraction aperture.  The left plots show the initial cross-section and phase space. The middle plots show the maximum and 2-rms envelopes for H$_2^+$. The vertical lines show the boundaries of the diagnostic boxes and magnets.  The right plots are the phase spaces at the end of the simulation. \label{fig:warp_beamline}}
\end{figure*}
Ions extracted from MIST-1 are simulated using two publicly-available software packages: IBSimu \cite{kalvas:ibsimu} and Warp \cite{friedman:warp, warp}.  IBSimu is used initially because it accurately models ions traversing the plasma sheath. However, IBSimu would be too computationally expensive to use for a simulation of the entire test beamline. On the other hand, Warp accurately models space-charge effects and beam transport, with less computing power than IBSimu, but the plasma sheath modeling in Warp is not well-established. This led to the decision to combine the packages in series. 
IBSimu simulates ions from the ion source to midway through the first 6-way cross, 
Warp then simulates ions from the first 6-way cross to the end of the beamline.

\subsection{Ion Source Extraction (IBSimu)}
The extraction system was designed and simulated using IBSimu, a particle-in-cell code developed at Yväskylä,  which uses iterative processes to calculate the particle trajectories through electromagnetic fields. 
IBSimu has successfully been used to design and simulate several extraction systems 
and is well-benchmarked against experimental results \cite{nishioka2018integrated, kalvas2015cw, toivanen2013double,midttun2017benchmarking}.
IBSimu uses electrode geometries that are imported from CAD files. The electrodes, set to static potentials, are then used to calculate the external fields in the system. The particle trajectories are simulated by ray-tracing of the particles through the superposition of the external field 
and the beam's self-field (space-charge). This process is repeated until the simulation converges.  Multiple species can be simulated simultaneously, accounting for the space-charge effects of each. IBSimu is used for the first 14 cm of beam line.

\subsection{Low Energy Beam Transport (Warp)}
Warp is a particle-in-cell Python package that has been developed since the 1980s at Lawrence Livermore National Laboratory and Lawrence Berkeley National Laboratory \cite{friedman:warp, warp}. The particle distributions from IBSimu simulations are loaded as the initial distributions for the Warp simulations. Particles are propagated using the wxy-slice package. The wxy-slice package calculates the transverse space-charge effects at each step, but ignores longitudinal space-charge effects, which is a good approximation for slow-changing DC beams, and brings significant simulation speed-up. 
Space-charge compensation (from residual gas ionization in regions free from electrostatic fields) is treated as a free parameter and simulated by modifying each species' current: $I_{comp} = I_i \cdot (1-f_e)$ 
with $f_e$ the space-charge compensation factor from literature \cite{reiser:beams}. A more accurate space-charge compensation model will be implemented in future simulations \cite{winklehner:scc}. Simulations are done for the entire beam line, and individual 
sets of parameters, matching emittance measurements and mass spectrometer measurements.

Included in the simulations are models of the magnetic fields and vacuum components.  In order to accurately model the quadrupole and dipole magnets, CAD models of the yokes and coils were imported to COMSOL \cite{comsol} to calculate 3d fields using finite
elements methods.  These fields are imported into Warp with a scaling factor to simulate different field strengths. Vacuum components are accounted for by using Warp's 
internal functions to generate conductor data from Boolean operations on primitive shapes.  
Conducting cylinders are used for the vacuum tubes and 6-way crosses; a rectangular box is used for the dipole chamber. Particles coming into contact with conductors are removed from the simulation and the currents are adjusted after each simulation step.  

At the beginning of the Warp simulations, the beam has a circular cross-section and nearly identical phase spaces for xx' and yy'. During simulations, the maximum and 2-rms beam envelopes are saved at each step and the phase space of all remaining particles is saved at the end of each simulation (see \figref{fig:warp_beamline} for a typical output plot).
The final currents are calculated by multiplying the initial currents by the fraction of particles transmitted for each species.
To determine the beam composition (species and currents) for the simulations, 
mass spectrum measurements were used.
By varying the dipole scaling factor in Warp, the current measured in FC2 during 
an actual dipole sweep can be simulated. Good mass separation is achieved by placing 
a 1~cm wide slit between dipole magnet and FC2 in the experiment as well as in the 
simulations. Emittance scans can be matched at the z position of the Allison 
scanners by creating a 2D histogram of the simulated
particle phase spaces in the same location.

The simulated dipole scans show good qualitative agreement with measured spectra.
The simulated phase spaces show excellent qualitative and good quantitative agreement with 
emittance scans using the Allison scanners. As an example, the experimental settings 
used to obtain the emittance scans shown in Section \ref{sec:emittance} were used 
in the described simulation deck and results are plotted in \figref{fig:warp_beamline}.
The species considered in this simulations are H$^+$, H$_2^+$, H$_3^+$, N$^+$, O$^+$, 
H$_2$O$^+$, N$_2^+$, and O$_2^+$, with ratios obtained from the respective  mass spectrum.
The higher mass ions are due to insufficient pumping time (air and water that had 
gotten into the source chamber during a filament change) and not due to the contamination
described in Section \ref{sec:contamination}. The space charge compensation factor $f_e$
was set to 0.5. The phase space plots on the right side of \figref{fig:warp_beamline}
should be compared to the emittance plots in \figref{fig:emi_meas}.

%% file: Sec4_Measurement.tex
\section{Measurements \label{sec:measurements}}

This section presents first measurements of ion species from the MIST-1 ion source.   This establishes the base level of performance for this device, which can be further tuned as needed for future applications.   Note that the ion source front plate has an exchangeable disc that lets us 
use different extraction aperture diameters. During the presented measurements, we used two diameters, 3~mm and 4~mm, hence
currents are usually reported as densities in mA/cm$^2$.

\subsection{Source Contamination}
\label{sec:contamination}

\begin{figure}[tb!]
    \includegraphics[width=\columnwidth]{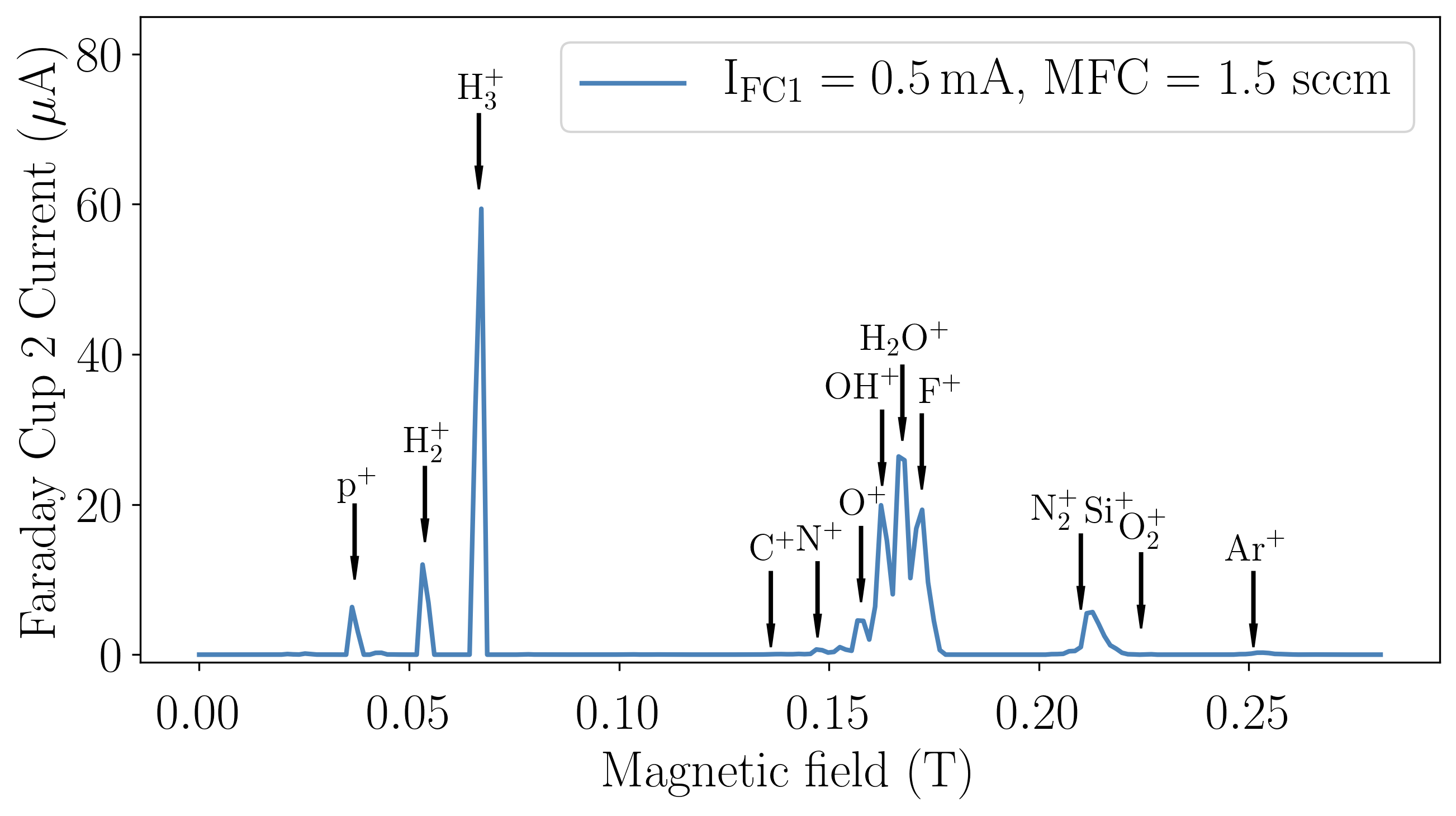}
    \caption{Mass spectrum with an $\mathrm{H}_2$ mass flow of 1.5 sccm.
             The most prominent species are indicated in the figure.
             Fluorine and Silicone most likely stem from the outgassing of 
             silicone-rubber O-rings. Nitrogen and water 
             (and the accompanying OH$^+$) can be
             attributed to a cracked ceramic insulator.
             \label{fig:contaminants}}
\end{figure}

The first systematic tests during early commissioning of the MIST-1 ion source 
showed a number of contaminants that persisted through all measurements.
As an example, \figref{fig:contaminants} shows a typical mass spectrum with
prominent species indicated.
Using M/Q for the mass-to-charge-state ratio in units of amu,
we have not found significant contamination above M/Q = 40 
(a small Ar peak can occasionally be seen) and restrict our dipole scans 
to M/Q = 0 to M/Q = 45 typically.
The M/Q of the peaks that were found in the recorded
spectra hinted at water contamination ($\mathrm{O}^+$, $\mathrm{OH}^+$,
$\mathrm{H}_2\mathrm{O}^+$),
a small air leak ($\mathrm{O}^+$, $\mathrm{O}_2^+$, $\mathrm{N}^+$, 
$\mathrm{N}_2^+$, $\mathrm{Ar}^+$), and 
a third source of contamination, related to outgassing of the silicone rubber
O-rings ($\mathrm{F}^+$, $\mathrm{Si}^+$).
Replacement of the back plate O-rings with Viton$^{\tiny{\textregistered}}$
and removal of a cracked ceramic insulator reduced the source contamination significantly.
Measurements that were taken prior to the O-ring replacement will be indicated as such.

\subsection{Performance Tests\label{sec:performance}}
The first results we are reporting here are of the ion source peak performance 
to date: The highest extracted current density, the highest \htp fraction, 
and the highest total extracted \htp current density (a balance between 
\htp fraction and total extracted current density).

\vspace{5pt}
\textbf{Highest total current.} With 5~mA of total beam current measured in Faraday cup 1, 
the highest current density recorded so far was $\approx 40$~mA/cm$^2$ 
(4~mm diameter aperture).
This was with a discharge voltage of 150~V and a \hyd flow of 1 sccm. Accordingly
(see discussion below) the species balance was shifted towards \hthp, with an
\htp fraction of $\approx 22$\,\% and \hthp fraction 
of $\approx 51$\,\%. This measurement was taken before O-ring replacement and, consequently,
suffered from contamination with higher mass species.

\begin{figure}[tb!]
    \includegraphics[width=\columnwidth]{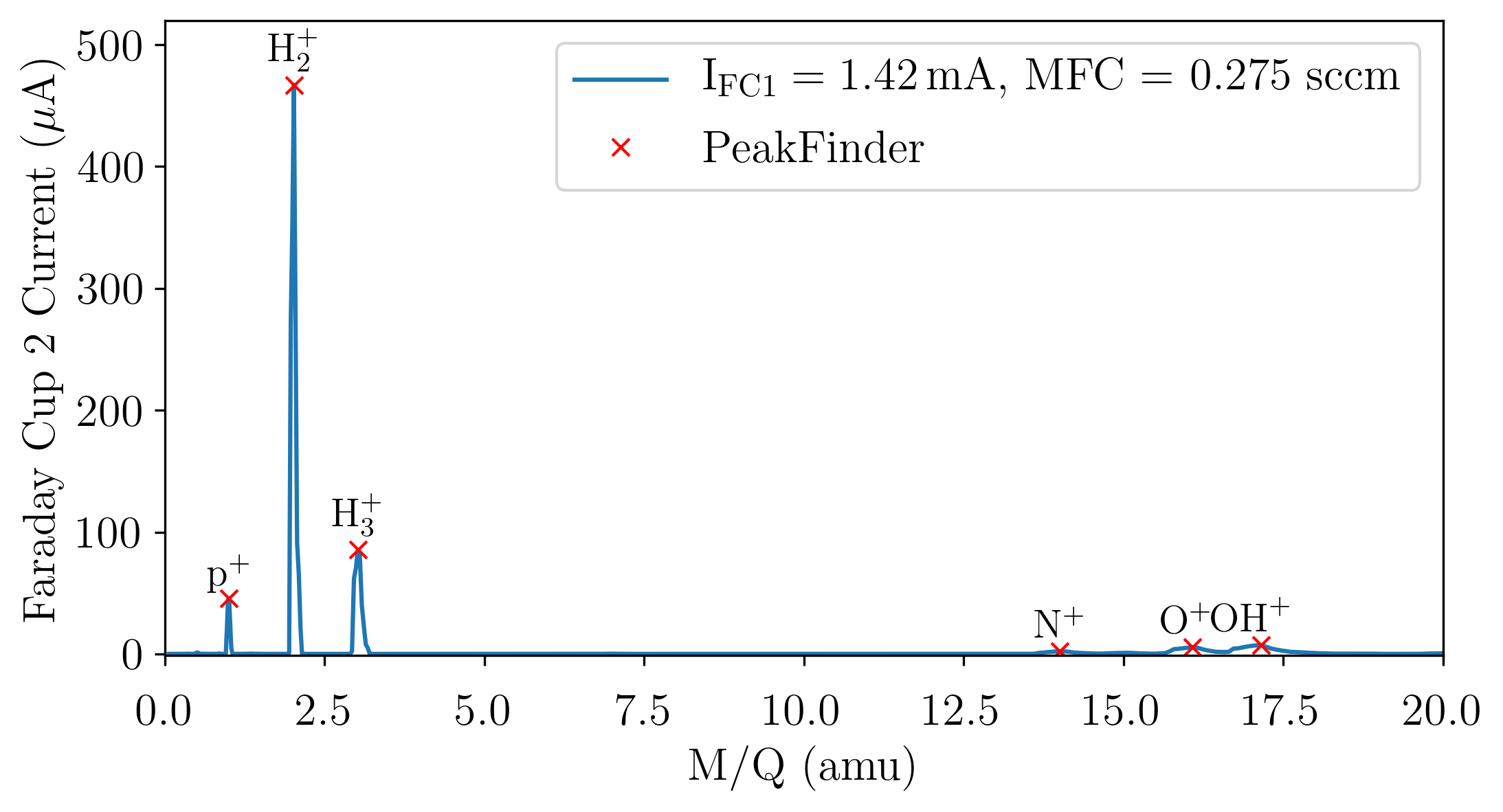}
    \caption{Mass spectrum for \hyd flow = 0.275~sccm and 125~V discharge voltage.
             The \htp fraction is 76\,\%. Note that contaminants are reduced to virtually 
             zero after O-ring replacement.
             \label{fig:high_h2p_2}}
\end{figure}

\vspace{5pt}
\textbf{Highest \htp contribution.} With a lower \hyd flow of 0.275 sccm, 
$\mathrm{V}_\mathrm{discharge} = 125$~V, 
and $\mathrm{I}_\mathrm{discharge} = 7$~A, the
highest fraction of \htp was recorded as 76\,\%. The total 
extracted current was 1.42 mA (11.4~mA/cm$^2$).
The corresponding mass spectrum is shown in \figref{fig:high_h2p_2}.
Coincidentally, this is also the highest current recorded with 
\htp being the dominant species. 470 $\mu$A were transported to the end of the beam
line, constituting a beam line transmission efficiency of 44\,\%, which agrees with Warp.

\subsection{Systematic Parameter Variations}
The ion source is designed to allow several parameters to be varied.  
Here we provide information on the total current and species
composition of the beam for each set of parameters.

\begin{figure}[b!]
    \includegraphics[width=\columnwidth]{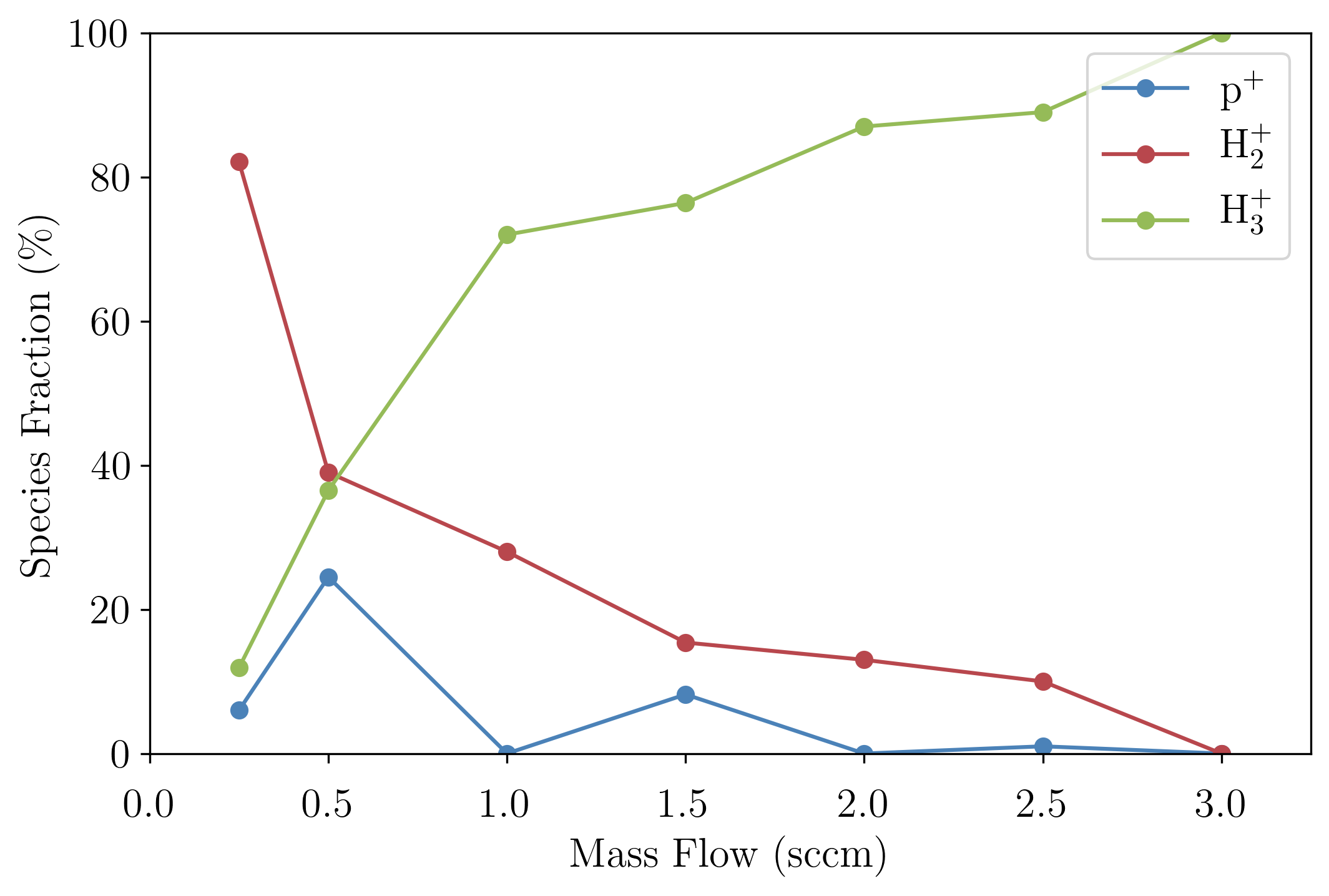}
    \caption{Variation of hydrogen flow from the MFC. This measurement 
             was taken prior to the replacement of O-rings and a constant 
             50\,\% contaminants are not included in the plot. 
             The typical trend of high \hthp with high mass flow can be observed.
             \label{fig:flowvar}}
\end{figure}

\vspace{5pt}
\textbf{Hydrogen flow.} In this study, the total extracted beam current was
held stable at 0.5~mA. The mass flow was varied from 0.25 to 3.0~sccm. The results are
plotted in \figref{fig:flowvar}. Notably, the \htp contribution rises towards 
lower pressures. This is quite typical, as with higher flow rates the 
probability for collisions of the already formed \htp ion with other
molecules, plasma ions, and electrons increases.
The contribution from the contaminant species (not shown in the plot) 
is nearly constant at $\approx50$\,\%. 
Throughout our measurements, the optimal mass flow for \htp production has been 
between 0.15 and 0.25~sccm.

\begin{figure}[tb!]
    \includegraphics[width=\columnwidth]{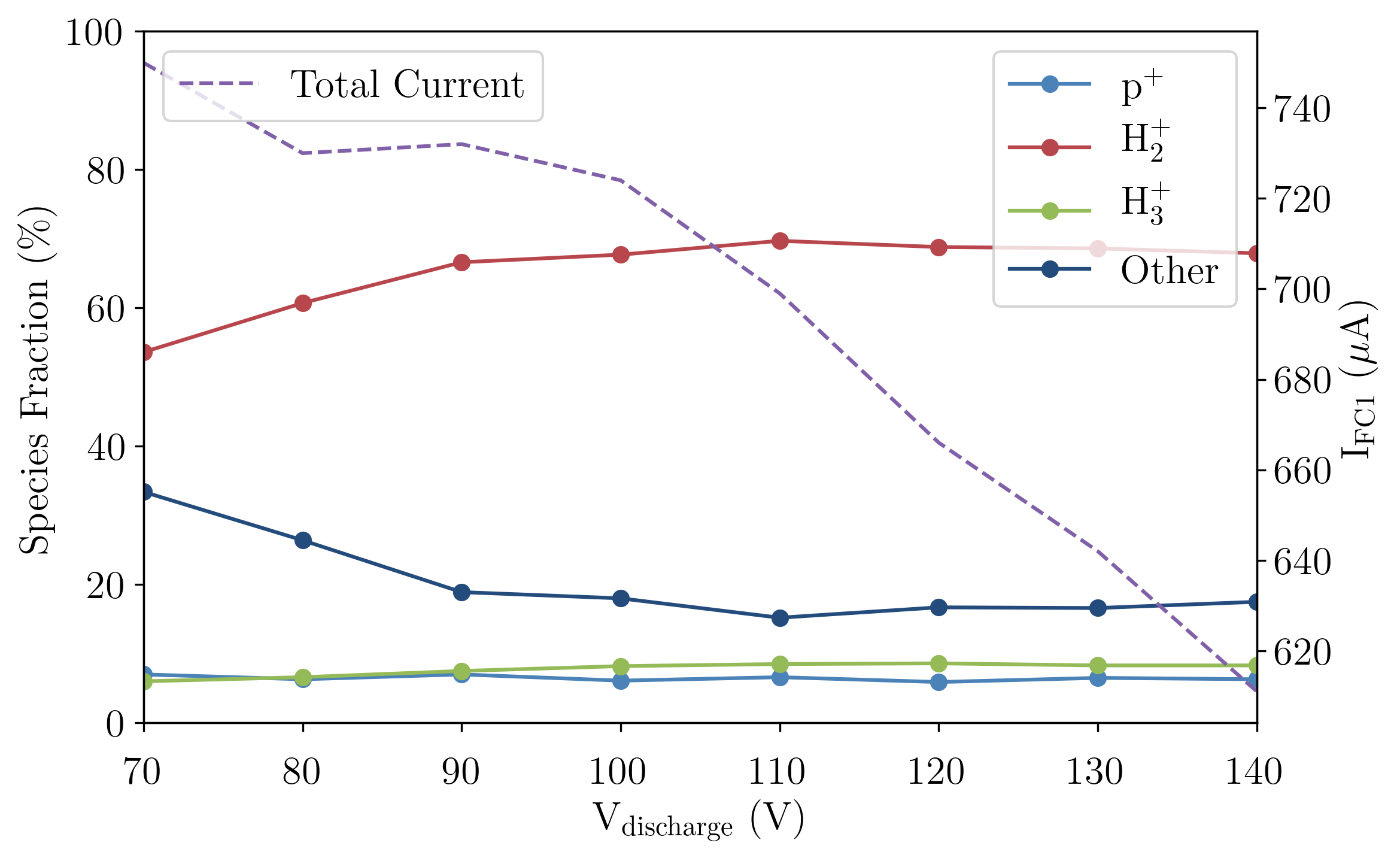}
    \caption{Variation of discharge voltage. Discharge current was kept constant at 
             4~A. To achieve this, the heating power of the filament is reduced with
             higher discharge voltage, which reduces the total extracted current 
             by about 120~$\mu$A (secondary axis).
             \label{fig:vdisvar}}
\end{figure}

\vspace{5pt}
\textbf{Discharge voltage.} Here we varied the discharge voltage from 70~V to
140~V, while keeping the \hyd flow rate constant at 0.2~sccm. The total extracted 
current in Faraday cup 1 was 0.75~mA and dropping.
This reduction stems from the adjustment of the filament heating current 
to keep the discharge current constant at 4~A throughout this measurement.
As can be seen in \figref{fig:vdisvar},
the contribution of the remaining contaminants is reduced, while \htp slightly increases
up to 110 V. After that, all contributions remain near constant.

\begin{figure}[t!]
    \includegraphics[width=\columnwidth]{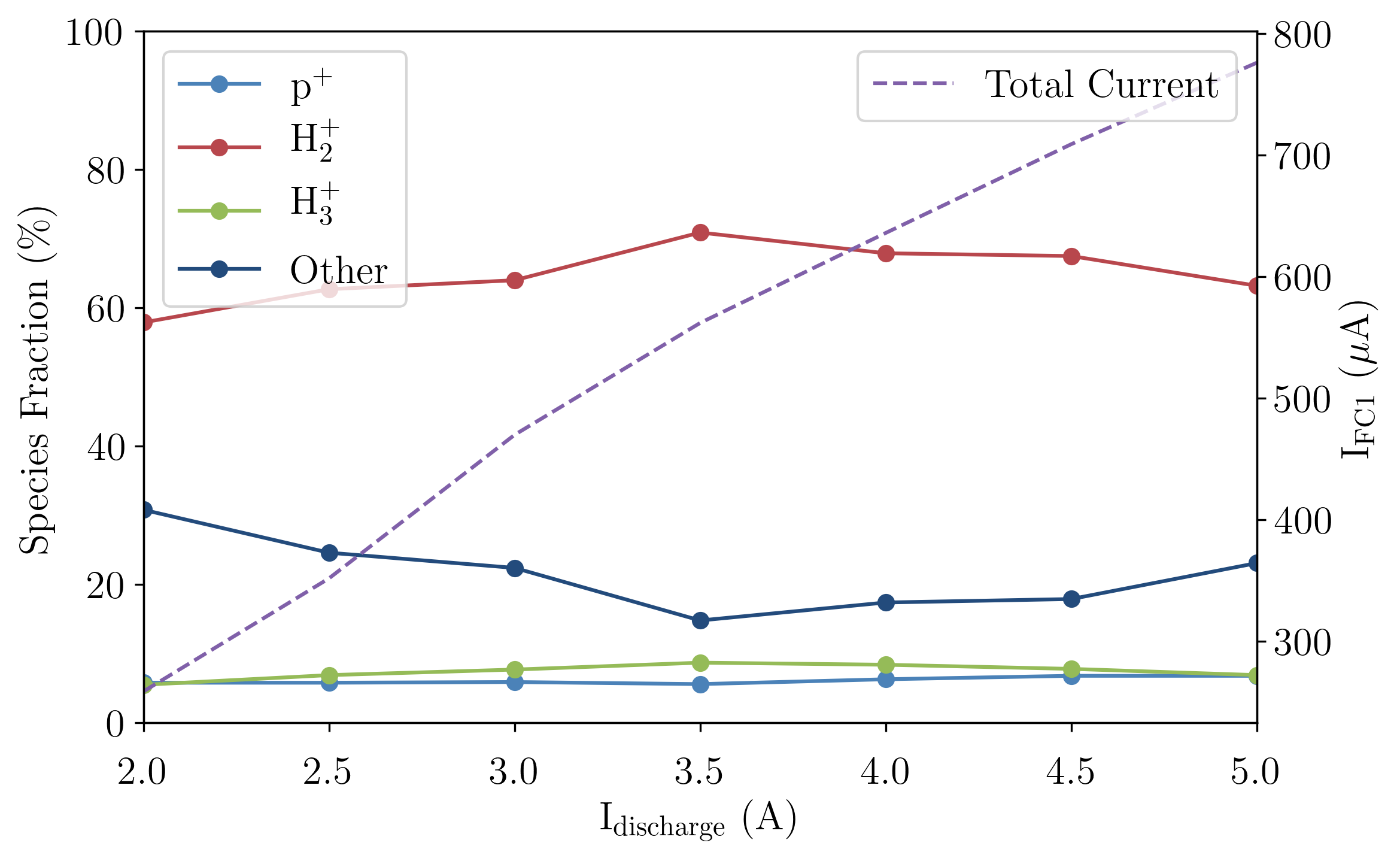}
    \caption{Variation of discharge current. Here the discharge voltage was kept at 120 V
             and \hyd flow at 0.2~sccm. No strong dependency is observed at these
             values. 
             \label{fig:idisvar}}
\end{figure}

\vspace{5pt}
\textbf{Discharge current.} Here we varied the discharge current from 2~A to
5~A, while keeping the \hyd flow rate constant at 0.2~sccm. While increasing the
discharge current, the total extracted current increased almost linearly. 
This is not surprising as a higher discharge current usually indicates higher 
plasma density. No strong dependence of the species ratios is observed in this
regime.

\subsection{Emittance Measurements}
\label{sec:emittance}
\begin{figure}[b!]
    \includegraphics[width=\columnwidth]{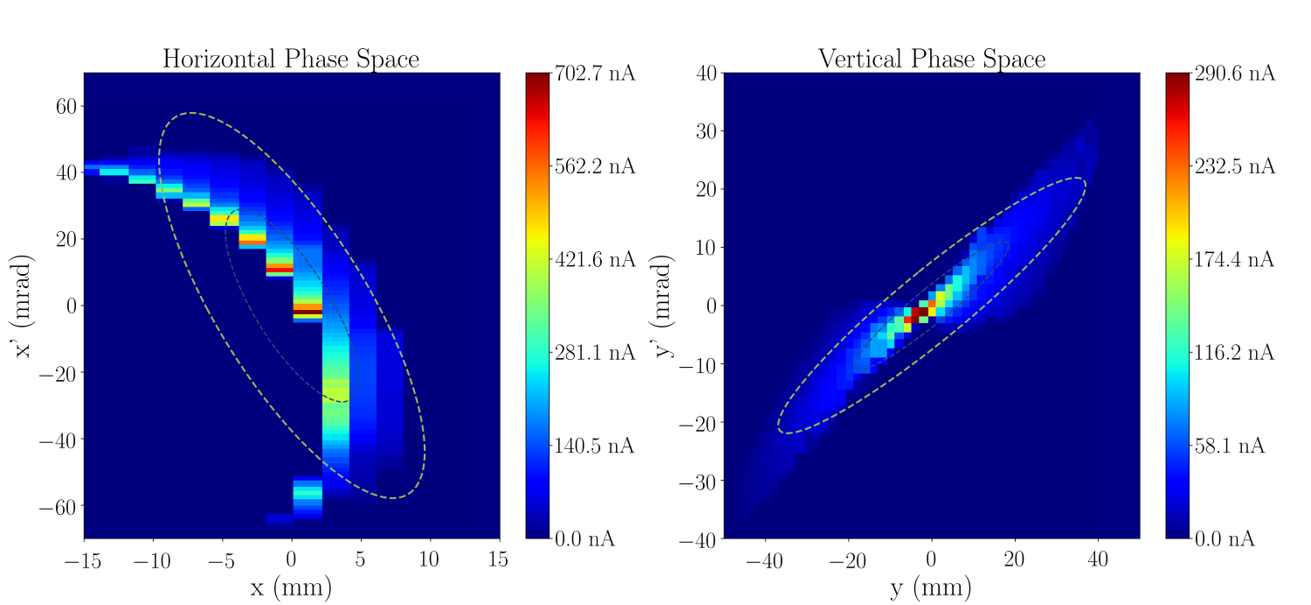}
    \caption{First emittance scans of a mass-separated \htp beam. 
             Good qualitative and quantitative agreement 
             was found with Warp simulations. Compare to \figref{fig:warp_beamline}, 
             right phase space plots.
             \label{fig:emi_meas}}
\end{figure}
An example of the results from the two Allison emittance scanners can be 
seen in \figref{fig:emi_meas}, for a mass-separated \htp beam, 
allowing a demonstration of good qualitative agreement with the IBSimu/Warp simulation 
of the setup (cf. Section \ref{sec:sims}). The measured 1-RMS, normalized emittances
at the end of the beam line were 0.37~$\pi$-mm-mrad and 0.29~$\pi$-mm-mrad for 
the horizontal and vertical scans, respectively. The total current
from the source was 850~$\mu$A and the \htp current arriving at the scanner
location was 200~$\mu$A.

\subsection{Discussion}
As \htp has a short mean
free path (3-5~cm) before it either combines with an H to \hthp or is dissociated into protons, 
the \hyd gas flow must be in balance with discharge voltage, and filament position,
with the filament position close to the extraction aperture being a critical 
aspect that is reflected in the design through the short source body.

\begin{figure}[t!]
    \includegraphics[width=\columnwidth]{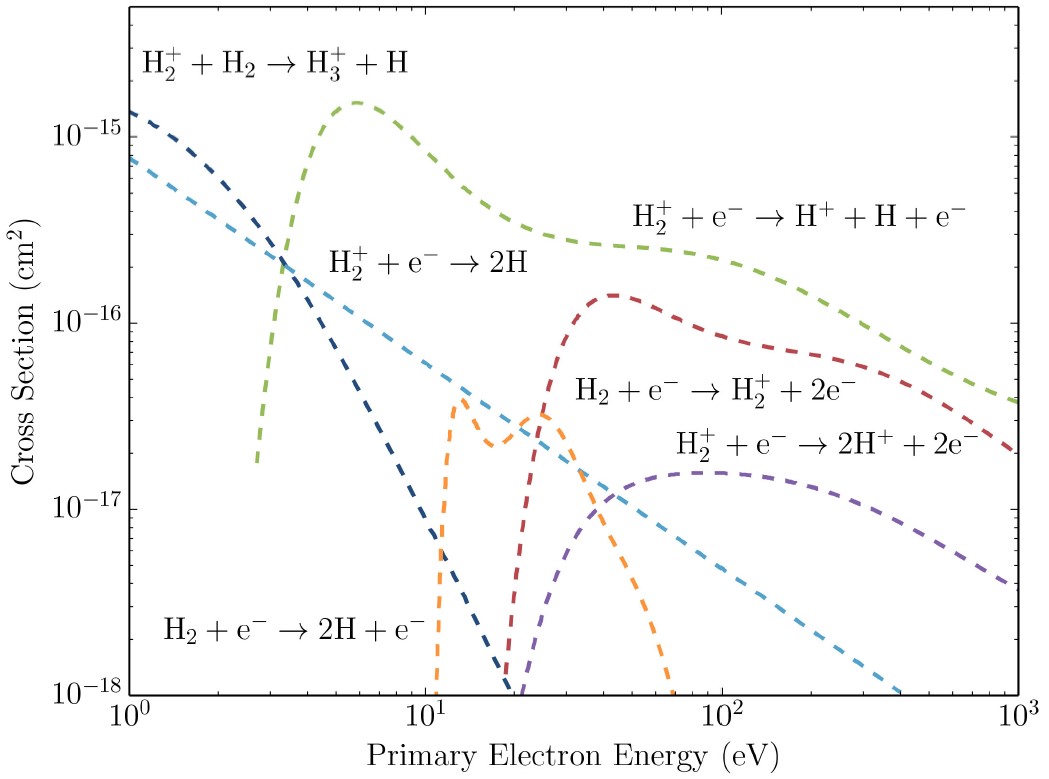}
    \caption{Rection cross-section for processes involved in the 
             production, recombination, and dissociation of \htp.
             From \cite{axani:mist2} \label{fig:h2p_xs}}
\end{figure}

The dominant process for \htp production is 
\begin{equation*}
    \hyd + \mathrm{e}^- \rightarrow \htp + 2 \mathrm{e}^-
\end{equation*}
with the three main processes reducing the available \htp being:
\begin{align*}
    \htp + \mathrm{e}^- & \rightarrow \hp + \mathrm{H} + \mathrm{e}^-, \\
    \htp + \hyd & \rightarrow \hthp + \mathrm{H}, \\
    \htp + \mathrm{e}^- & \rightarrow 2 \hp + 2 \mathrm{e}^-.
\end{align*}
The corresponding cross-sections are plotted in \figref{fig:h2p_xs}.
Our measurements indicate that \htp becomes the dominant 
species at low \hyd mass flow (0.25 sccm and below) and low discharge voltage 
(80~V to 120~V). 
This is also in agreement with earlier findings by Ehlers and 
Leung \cite{ehlers:multicusp1, ehlers:multicusp2}.
In the next measurement period, the filament position will be varied to
find the maximum extracted current with high \htp fraction.
Other planned improvements to the source include better water cooling and
a plasma aperture made from a tungsten alloy (75\% W, 25\% Cu)
to improve heat transfer away from the source aperture and to provide 
better thermal stability.

The quality (emittance) of beam extracted from the ion source is very good, 
as is typical for filament-driven multicusp ion sources. Currently, we only
have emittance scans after the beam has gone through the non-ideal LEBT, but 
as the agreement with the simulations is good, we can extrapolate back to the
ion source, which yields normalized 1-rms emittances below 0.05~$\pi$-mm-mrad 
at extraction. 
The large emittance growth in the LEBT can be attributed to the strong 
multipole components in the dipole fringe fields. Borrowed equipment was used
to assemble the test stand. Furthermore, the dipole magnet has no vertical focusing
and a pair of quadrupole magnets were used to compensate for this.
In a future measurement, the Allison scanners will be moved into
DB1 to measure the combined emittance of the beam (at $>80$\,\% \htp) right after 
extraction, which will decouple the emittance measurement from the 
shortcomings of the LEBT, albeit at the expense of having some proton 
and \hthp contamination in the scans.

Contamination of the source with heavier-mass species in 
a previous commissioning run was significantly reduced by replacing a 
pair of O-rings and a cracked ceramic insulator. Once the source has warmed
up, contaminants are virtually non-existent. 
The source now operates stably over periods of several hours without significant 
changes in beam current and composition.

As a final point of discussion, we would like to mention that the stainless steel
body of the source, while robust, turned out not to be ideal in terms of
transferring heat away through the water cooling channels. Particularly, the 
plasma aperture shows distortions after long running times (several hours) at high power.
We are beginning to replace individual source parts with a high yield-strength 
copper alloy like Elmedur.

%% file: Sec5_Conclusion.tex
\section{Outlook and Conclusion}

We have presented a new filament-driven multicusp ion source, 
designed to produce high currents of \htp in DC mode for long-time operation, and 
established simulations that provide good qualitative agreement with the 
constructed device.
We reported maximum currents of 1.42~mA from a 4~mm aperture consisting of 
76\,\% \htp, which corresponds to a current of 1.08 ~mA of \htp. To our knowledge,
this is the highest \htp current published to date from a multicusp ion source in 
DC mode, while \htp was the dominant species. This is an important milestone for
MIST-1 and, as described below, we have a clear path for further improvement.

For the novel RFQ Direct Injection method that is being proposed for the IsoDAR 
experiment \cite{winklehner:rfq1, winklehner:rfq2, winklehner:nima}, 
the nominal goal is 10 mA of \htp delivered by the ion source.
If our 4 mm diameter aperture results are scaled up to an 8 mm aperture 
(assuming constant plasma density), 4.3~mA of \htp can currently be delivered 
from this source. This is only a factor 2.5 short of the goal. 
With the ongoing systematic tests of filament material, shape, and position as 
well as a stronger magnetic confinement field, the available \htp beam current 
will be significantly increased.
Furthermore, improved cooling will allow higher discharge currents, also leading to 
higher plasma densities and thus, higher currents.
Even with the currently reachable 4.3~mA of \htp, a compact cyclotron of the IsoDAR
design could be built that delivers 4 times more proton current (2~mA of \htp
coresponds to 4~mA of protons after charge-stripping) than commercially available 
machines, and also beats record holder PSI Injector II, which 
delivers a maximum of 2.7~mA protons \cite{seidel:psi_status}.

In addition to construction of the source, we developed an accurate simulation
model of the ion source and beam line, that was compared with mass spectra and 
emittance measurements in the second 6-way cross (DB2) with good agreement. 
Thus, although the LEBT itself has certain shortcomings (due to the use of borrowed 
equipment), we understand the beam dynamics well and can extrapolate the beam quality 
directly after the source from the measurements at the end of the LEBT. The extrapolated
emittance at extraction is $<0.05$ $\pi$-mm-mrad (1-rms, normalized), which is well below
the IsoDAR requirements.